\documentclass[12pt,preprint]{aastex}

\usepackage{psfig,natbib}   
\usepackage{emulateapj5}   

\shorttitle{The host galaxy of GRB\,990705}
\shortauthors{E.\,Le Floc'h et al.}

\slugcomment{Received 2002 September 6; Accepted 2002 November 12}

\begin{document}

\title{VLT and HST observations of the host galaxy of GRB\,990705}

\def\SAp{3}
\def\IAFE{4}
\def\IFA{5}
\def\MAX{6}
\def\LAPLATA{7}
\def\ISDC{8}
\def\GEN{9}
\def\IFC{10}
\def\ESO{11}

\author{E.~Le~Floc'h\altaffilmark{\SAp},
P.-A.~Duc\altaffilmark{\SAp},
I.F.~Mirabel\altaffilmark{\SAp,\IAFE},
D.B.~Sanders\altaffilmark{\IFA,\MAX},
G.~Bosch\altaffilmark{\LAPLATA},
I.~Rodrigues\altaffilmark{\SAp},
T.J.-L.~Courvoisier\altaffilmark{\ISDC,\GEN},
S.~Mereghetti\altaffilmark{\IFC},
\&\
J.~Melnick\altaffilmark{\ESO}}

\altaffiltext{1}{Based on
       observations with the Very Large Telescope, obtained at the
       European Southern Observatory in Chile under Prop. 68.B-0250(B)}
\altaffiltext{2}{Based on
       observations with the NASA/ESA Hubble Space Telescope, obtained
       at the Space Telescope Science Institute, which is operated by
       the Association of Universities for Research in Astronomy,
       Inc.~under NASA contract No.~NAS5-26555.}
\altaffiltext{3}{CEA/DSM/DAPNIA Service d'Astrophysique, F-91191 Gif-sur-Yvette, France (elefloch, paduc, fmirabel, irapuan@cea.fr)} 
\altaffiltext{4}{Instituto de Astronom\'\i a y F\'\i sica del Espacio, cc 67, suc 28. 1428 Buenos Aires, Argentina} 
\altaffiltext{5}{Institute for Astrononmy, University of Hawa\"{\i}, 2680 Woodlawn Drive, Honolulu, HI 96822, United States (sanders@IfA.Hawaii.Edu)}
\altaffiltext{6}{Max-Planck Institute for Extraterrestriche Physik, D-85740, Garching, Germany}
\altaffiltext{7}{Facultad de Cs. Astronomicas y Geofisica, Paseo del Bosque s/n, La Plata,
Argentina (guille@lahuan.fcaglp.unlp.edu.ar)}
\altaffiltext{8}{INTEGRAL Science Data center, Ch. d'Ecogia 16, CH-1290 Versoix, Switzerland (Thierry.Courvoisier@obs.unige.ch)}
\altaffiltext{9}{Geneva Observatory, Ch. des Maillettes 11, 1290 Sauverny, Switzerland}
\altaffiltext{10}{Istituto di Astrofisica Spaziale e Fisica Cosmica, Sezione di Milano "G. Occhialini," via Bassini 15, I-20133 Milan, Italy (sandro@mi.iasf.cnr.it)}
\altaffiltext{11}{European Southern Observatory, Alonso de Cordova 3107, Santiago, Chile (jmelnick@eso.org)}

\begin{abstract}
We present VLT spectroscopic observations of Gamma-Ray Burst
(GRB)\,~990705 host galaxy and highlight the benefits provided by the
prompt-phase features of GRBs to derive the redshifts of the latter.
In the host spectrum, we indeed detect an emission feature which
we attribute to the [OII]\,$\lambda\lambda$\,3726/3729\,\AA \, doublet
and derive an unambiguous redshift z=0.8424\,+/--\,0.0002 for this
galaxy. This is in full agreement with the value
z$\sim$0.86\,+/--\,0.17 previously derived using a transient
absorption edge discovered in the X-ray spectrum of GRB\,990705.  This
burst is therefore the first GRB for which a reliable redshift was
derived {\it from the prompt phase emission itself\,}, as opposed to
redshift determinations performed using putative host galaxy emission
lines or intestellar absorption lines in the GRB afterglows.  Deep and
high resolution images of the host of GRB\,990705 with the HST/STIS
camera reveal that the burst occured in a nearly face-on Sc spiral
galaxy typical of disk-dominated systems at
0.75\,$\leq$\,z\,$\leq$\,1.  Assuming a cosmology with
H$_0$\,=\,65~km~s$^{-1}$\,Mpc$^{-1}$, $\Omega_m$\,=\,0.3 and
$\Omega_{\lambda}\,=\,0.7$, we derive an absolute B magnitude 
M$_B$\,=\,--21.75 for this galaxy and a star formation rate
SFR~$\approx$~5--8~M$_{\odot}$\,yr$^{-1}$.  Finally, we discuss the
implications of using X-ray transient features to derive GRB redshifts
with larger burst samples, and especially examine the case of short
and dark-long GRBs.

\end{abstract}


\section{Introduction}  

Since the discovery of their X-ray, optical and radio transient
counterparts, the cosmic Gamma-ray bursts (GRBs) have been regarded as
one of the most promising tools to probe the star formation in the
early Universe \citep{Totani97,Wijers98,Mirabel00,Blain00}.  There is
indeed increasing evidence that the long and soft GRBs originate from
the core collapse of massive stars within starburst regions of distant
galaxies \citep[e.g.,][]{Bloom02a}. Since they are likely detectable up
to very high redshifts \citep{Lamb00}, GRBs could soon open a new
window to sample the star-forming activity at cosmological lookback
times, and ultimately provide a new glimpse on galaxy evolution.

The possibility to detect emission and/or absorption features in the
spectra of GRBs and their afterglows is among the most outstanding
benefits of the high--z galaxy selection by these events.  Such
detections can indeed be done independently of the GRB host
luminosities and have already enabled spectroscopic redshifts of very
faint galaxies to be derived \citep[e.g.,][]{Vreeswijk01}. This
perspective strongly contrasts with the deep survey observations which
can only provide photometric redshifts for the faintest sources.
Nonetheless, the correct GRB redshift identifications from the lines
detected in {\it afterglow} spectra are not always straightforward.
Absorption features observed in the optical continuum of GRB
counterparts may indeed originate from foreground absorbers
\citep[e.g.,][]{Metzger97}, while the interpretation of the emission lines
detected in the X-ray afterglows has already led to some
mis-identifications of host redshifts.
\citep[i.e., GRB\,970828:][]{Yoshida99,Djorgovski01a}.

In this context, the GRB\,990705 event is of remarkable interest.
Using data from the BeppoSAX satellite, \citet{Amati00} reported the
discovery of a transient absorption edge at $\sim$\,3.8~keV in the
prompt X-ray emission of this burst, which they interpreted as the
GRB-intrinsic signature of an iron-enriched absorbing medium at
z\,$\sim$\,0.86 \citep[see also][]{Lazzati01}.  This has been so far
the only GRB for which a feature allowing a possible redshift
determination was observed during the {\it prompt emission} of the
burst itself.  Following this event, an optical and near-infrared
follow-up by \citet{Masetti00} led to the discovery of a red and
rapidly-decaying afterglow localized behind the Large Magellanic Cloud
(LMC), while deep optical images of the burst location performed with
the HST \citep{Holland00c} and the VLT \citep{Saracco01b} revealed an
R\,$\sim$\,22.2\,mag underlying host galaxy.

In this letter we report on VLT observations carried out to derive the
spectroscopic redshift of this host galaxy, and which allowed us to confirm
the redshift of GRB\,990705 derived by \citet{Amati00}.  We also
analyse  public HST data of the host.

\section{Observations and data reduction}  

The spectroscopic observations of the GRB\,990705 host galaxy were
performed on 2001 December 21 and 22 (burst 
trigger\,+ $\sim$\,900\,days) with the FORS2 instrument installed on
the VLT UT4/Yepun at ESO. Spectra were obtained under moderate seeing
conditions ($\sim$ 1\,{\arcsec}) using a medium resolution grism
(600RI) in combination with a 1\,{\arcsec}-width slit, and
totalizing an integration time of 1.5\,hours. We thus covered
an effective wavelength range $\sim$~5600 -- 8000\,\AA \, 
with an instrumental resolution $\sim$\,4.5\,\AA. The slit was positioned on
the sky so as to cover the outer region of the galaxy where the burst
occured.
The galaxy spectra were flux-calibrated
using spectroscopic standard stars.

The HST observations of the GRB\,990705 host galaxy\footnote{see also
http://www.ifa.au.dk/$\sim$hst/grb\_hosts/data/grb990705/index.html}
were taken and reduced by  \citet{Holland00c,Holland00a} 
as part of the ``Survey of the Host Galaxies of Gamma-Ray
Bursts'' using the STIS camera.  Images were
obtained with the 50CCD (clear, pivot $\lambda_o$\,=\,5835\,\AA,
hereafter CL) and F28X50LP (long pass, pivot
$\lambda_o$\,=\,7208\,\AA, hereafter LP) apertures, respectively on
2000 July 25 and 2000 August 25 (i.e., $\sim$\,400 days after the
burst).  The respective total exposure times were 8851\,s and 8202\,s
in the CL and LP apertures.  
We deconvolved the
images following a multi-resolution wavelet decomposition
\citep{Starck98} and the use of PSFs obtained from the combination
of foreground stars in the images.

The photometry measurements were performed on the data before
deconvolution to preserve reliable flux and noise estimates.  We
corrected the CL and LP aperture data from absorptions
A$_{CL}$\,=\,0.36 and A$_{LP}$\,=\,0.28 mag assuming the
extinction curve of \citet{Cardelli89} and  the Galactic\,+\,LMC
extinction $E(B-V)$\,=\,0.12 obtained by 
 \citet{Dutra01}. Moreover, we carried out a careful
analysis using a multi-resolution transform method to subtract from
the images the multiple LMC foreground stars superimposed on the plane
of the galaxy.

\label{sec:red}  

\section{Results}
\label{sec:results}

\subsection{The redshift of the GRB\,990705 host galaxy}

The final VLT spectrum is shown in Figure\,1.  An emission feature is
clearly detected at $\sim$\,6868\,\AA \, in a region of the spectrum
where the residuals from the sky line subtraction are negligible.
Attributing this feature respectively to H$\alpha$ or Ly$\alpha$ would
imply redshifts z\,=\,0.05 and z\,=\,4.66, which is inconsistent with
the spiral morphology and the angular size of the galaxy (see section
3.2).  The line can thus only be due to
[OII]\,$\lambda\lambda$\,3726/3729\,\AA. It is actually not resolved
in our spectrum, but its width is in fact consistent with that of the
[OII] doublet.  Note that the low signal to noise ratio longward of
7300\,\AA \, does not allow us to detect H$\delta$ and H$\gamma$.
From the [OII] line, we derive a secure heliocentric redshift
z\,=\,0.8424\,+/--\,0.0002 ~for the host galaxy and GRB\,990705. This
is in full agreement with the value z\,$\sim$\,0.86\,+/--\,0.17
obtained by \citet{Amati00} from the transient feature observed in the
GRB prompt emission, and also appears consistent with the redshift
z\,=\,0.843 already mentioned by \citet{Lazzati01}. Assuming a
standard cosmology with H$_0$\,=\,65~km~s$^{-1}$\,Mpc$^{-1}$,
$\Omega_m$\,=\,0.3 and $\Omega_{\lambda}\,=\,0.7$, we thus measure for
the host of GRB\,990705 a luminosity distance d$_l$\,=\,5.8\,Gpc, and
a projected scale of 8.2~proper kpc (or 15.2~comoving kpc) per
arcsecond on the sky.  Because of the extended and rather diffuse
emission of the galaxy (see section\,3.2), we did not obtain a secure
estimate of the [OII] integrated flux lying outside of the slit, and
thus we could not derive its [OII] total luminosity. We roughly
measured, though with large uncertainties, an observed [OII]
equivalent width EW\,$\approx$\,40\,\AA, i.e., $\approx$\,20\,\AA \,
in the rest frame.

\begin{minipage}[b]{8.5cm}
\vskip 0.5cm
\centerline{\psfig{file=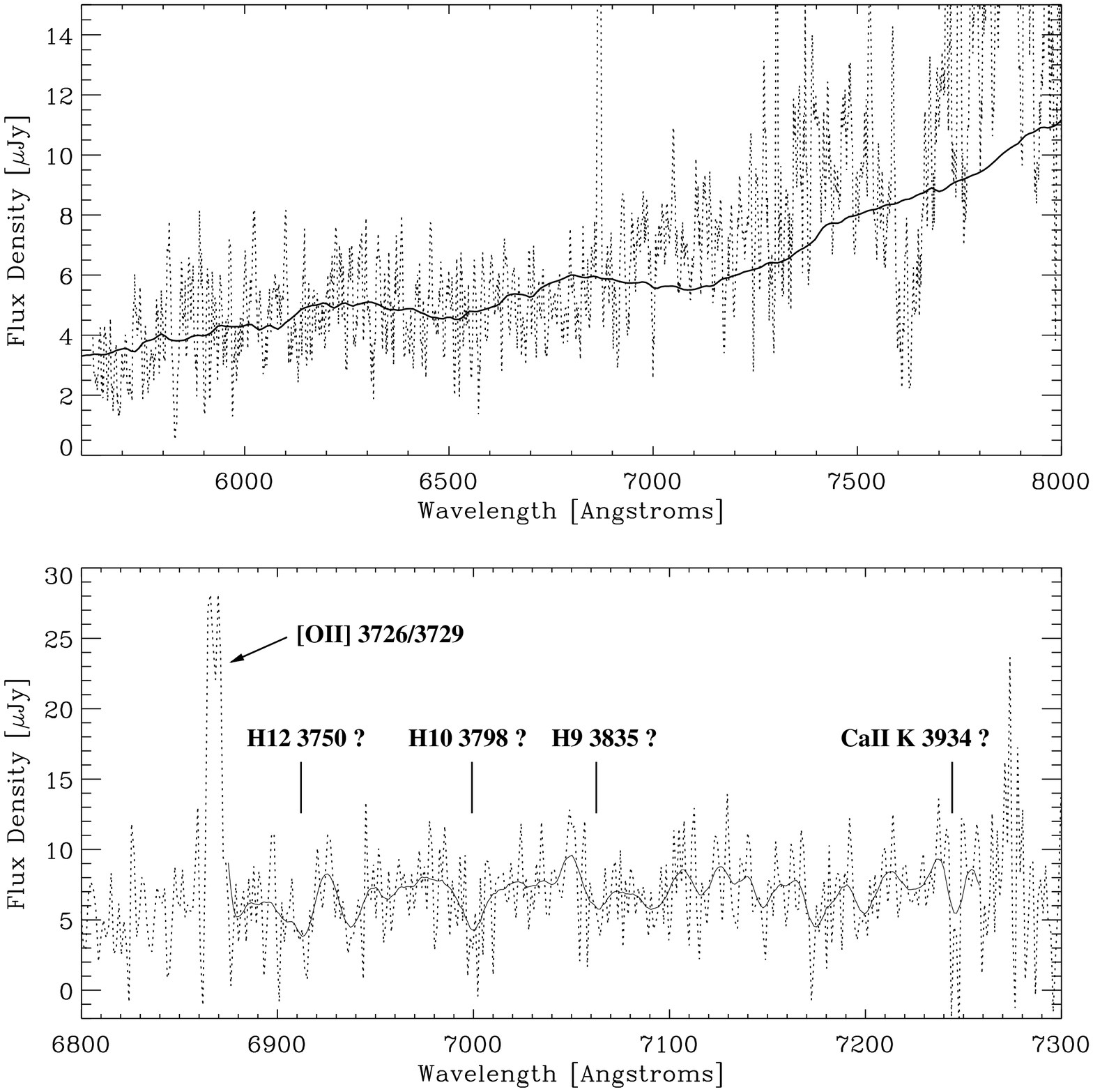,width=8.5cm,angle=0}}
\vskip 0.2cm \figcaption{{\it Top:\,} The dereddened VLT spectrum
smoothed by 5 pixels (dotted line), overlaid with the template of a
local Sc galaxy shifted to z\,=\,0.8424 (solid line). {\it Bottom:\,}
A zoom in the $\sim$ 6800--7300\,\AA \, range (dotted line), with the
smoothed continuum (solid line) superimposed to emphasize on the
possible detection of several stellar absorption lines as indicated in
the panel.}
\label{fig:sed}
\end{minipage}
\vskip .3cm

In addition to the [OII] emission doublet, we tentatively
detect several stellar absorption features at a similar redshift such
as H12 ($\lambda$\,3750\,\AA), H10 ($\lambda$\,3798\,\AA), H9
($\lambda$\,3835\,\AA) and CaII\,K ($\lambda$\,3934\,\AA). 
The reliabilty of these features is yet questionable given the low signal to
noise ratio in the continuum.
Finally, in spite of the poor sky subtraction longward of 7300\,\AA,
there is a hint for a break in the continuum around 7400\,\AA \, (see
Figure\,1).  This suggests the presence of the rest-frame
4000\,\AA-break as indicated by the comparison of our spectrum with
the template of an Sc galaxy \citep{Mannucci01} shifted to z\,=\,0.84, and provides
further support to our redshift determination.

\subsection{Structural parameters of the galaxy}

High resolution HST images reveal that the host
 is a face-on Sc spiral galaxy.
In Figure\,2 we show a pseudo true-color image of that source,
constructed by registering the deconvolved CL and LP
data. Two primary spiral arms following a ``m=2'' wave density mode
clearly extend from the northern and southern sides of the central
bulge, while other secondary arms are also observable.  The disk
spreads over a region of $\sim$\,3'' in diameter, with a half-light
radius R$_{\,0.5}$\,$\sim$\,7.5\,kpc (0.9\,{\arcsec}) in the range of
those characterizing the disk-dominated galaxies at
0.75\,$\leq$\,z\,$\leq$\,1 \citep{Lilly98}.

After subtracting the foreground stars, we performed photometry within
a 2''-radius aperture centered on the host nucleus using the STIS zero
points of \citet{Gardner00}. Dereddened AB magnitudes
M$_{AB}$(CL)\,=\,22.45\,+/--\,0.10
and M$_{AB}$(LP)\,=\,22.0\,+/--\,0.1 
were respectively derived from the CL and LP images, the uncertainties
being dominated by the subtraction residuals of the stars projected
ahead of the galaxy.

\begin{minipage}[b]{8.5cm}
\vskip .3cm 
\centerline{\psfig{file=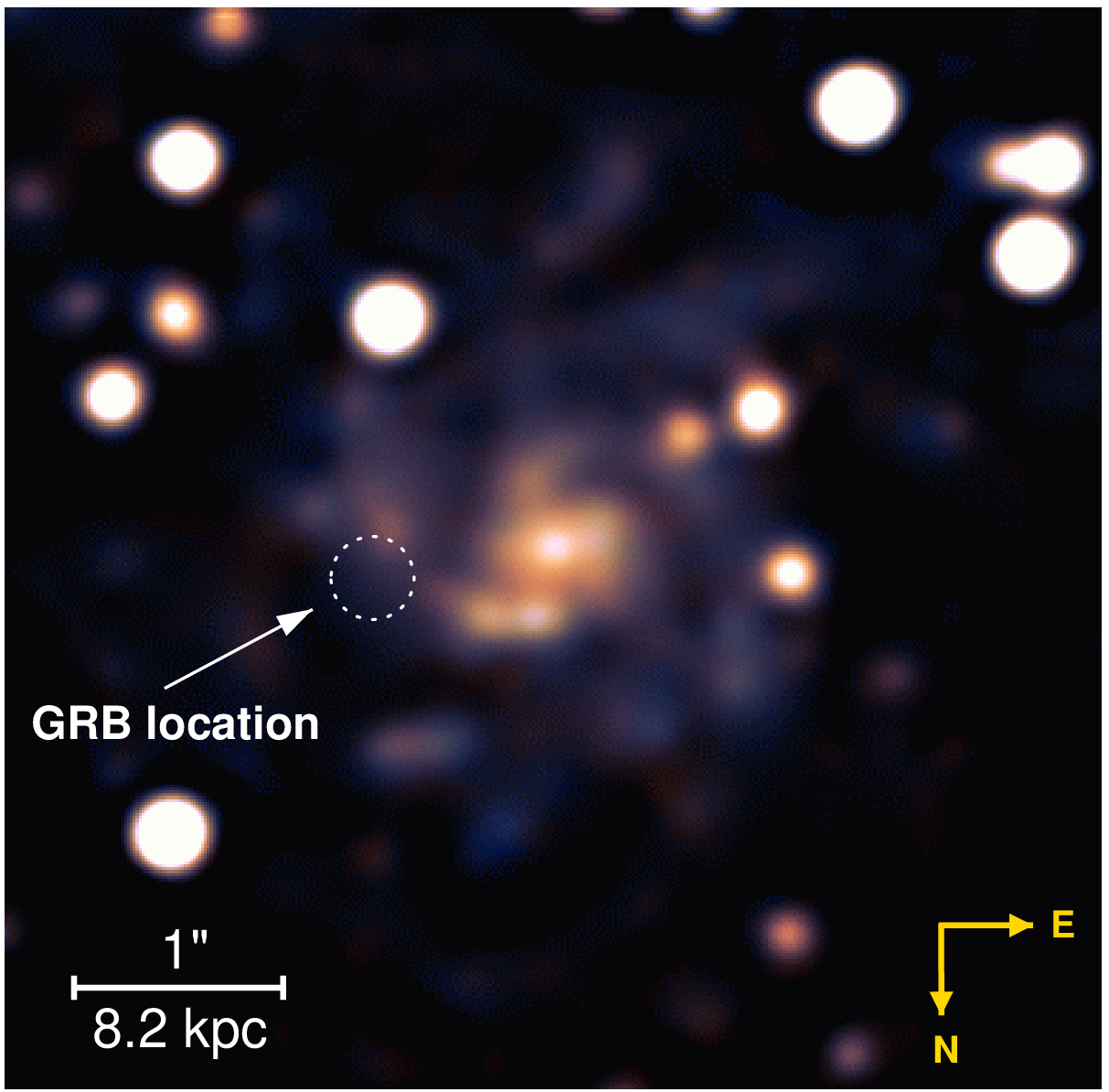,width=8.5cm,angle=0}}
\vskip 0.2cm \figcaption{A pseudo true-color image of the GRB\,990705
host galaxy, resulting from the combination of the drizzled and
deconvolved images taken through the CL and LP apertures.  The
3\,$\sigma$--error location of GRB\,990705 event as derived by
\citet{Bloom02a} is indicated by the dashed circle. Note that the GRB
occured in the outskirts of a star forming region within one of the
primary spiral arms.  {\it See the electronic edition of the Journal
for a color version of this figure.}}
\vskip .2cm
\label{fig:hst}
\end{minipage}

An attempt was made to fit the surface brightness profile of the
galaxy using a "bulge + exponential disk" decomposition.
Disentangling between the two contributions in distant sources requires
not only a proper correction from the PSF effects
\citep[e.g.,][]{Moth02} but also a sufficient sensitivity up to
$\sim$\,10 to 15~kpc in the averaged profile of the disk to reliably
constrain the scale length of the exponential component \citep[see
e.g., Fig.\,1 of][]{Rigopoulou02}. Because of the diffuse and extended
emission of the galaxy and the presence of the foreground stars, this
was hardly achieved in our data.
With large uncertainties, we suggest however that the host
galaxy is dominated by a large exponential disk (scale length
$\sim$\,5--6~kpc), with a rather small bulge contribution leading to
a bulge-to-total light ratio B/T $\approx$~0.10--0.15.  We also
estimated the central surface brightness of the disk component in the
dereddened LP data $\mu_0$(LP), which was then converted into a
rest-frame B$_{AB}$ magnitude by applying a cosmological dimming term
and a {\it k}-correction color factor as follows\,:

\begin{displaymath}
\mu_0(B_{AB}) = \mu_0(LP_{AB}) - 2.5 \log(1+z)^3 + (B_{obs}-LP)_{AB}  
\end{displaymath}

Since the LP aperture samples the almost rest-frame B~emission at the redshift
of the host, the color term should be rather small.  Using the
spectral energy distribution (SED) of a local Sc spiral galaxy shifted
to z\,=\,0.84 (see section\,3.3), we derive a {\it
k}-correction $\sim$\,0.1~mag and finally obtain a value
$\mu_0$(B$_{AB}$)~$\approx$~20.8. This is actually less than the canonical
\citet{Freeman70} value $\mu_0$(B$_{AB}$)=21.6 observed in local
disks, but is consistent with those found in higher-z spirals
\citep{Lilly98}. It is therefore in agreement with the observed global
trend for the disk central surface brightness to significantly
increase with redshift up to $\sim$~1 \citep{Lilly98}.

\subsection{Absolute B magnitude and star formation rate}

At z=0.84, the {\it rest-frame} B emission of the galaxy is shifted to
$\sim$\,8100\,\AA.
The dereddened 
continuum of the host of GRB\,990705 and the CL \--- LP color derived
from the STIS images are however consistent with the SED of a local Sc
spiral galaxy \citep{Mannucci01} shifted to z=0.84 assuming no
evolution.  To estimate the luminosity at B$_{rest}$, we thus
extrapolated the continuum of our spectrum using the template of
\citet{Mannucci01} and found a flux density
F$_{\nu}$\,($\lambda$\,=\,8100\,\AA)\,$\approx$\,12\,$\mu$Jy. Given
the assumed cosmology and the luminosity distance of the galaxy, this
implies a rest-frame absolute B-band magnitude M$_B\,\approx$\,--21.75
corresponding to a 2\,L$_*$ galaxy at z$\sim$1 \citep{Lilly95}. 

Using a similar method, we also measured the continuum luminosity at
$\lambda_{rest}$\,=\,2800\,\AA \, to estimate the level of
UV--unobscured star formation activity. From the Sc
template SED, we estimate a flux $\sim$~1.8\,$\mu$Jy at the
corresponding $\lambda_{obs}$\,=\,5152\,\AA \, and deduce a UV
luminosity L$_{UV} \sim$~4~erg~s$^{-1}$\,Hz$^{-1}$ at the redshift of
the host. Following the calibration of \citet{Madau98} and assuming a
\citet{Salpeter55} or \citet{Scalo86} Initial Mass Function, we
finally derive star formation rates
SFR~$\approx$~5~M$_{\odot}$\,yr$^{-1}$ or
SFR~$\approx$~8~M$_{\odot}$\,yr$^{-1}$, which are fairly common values
for star-forming galaxies at z$\sim$1 \citep{Lilly95}.

\section{Discussion and Conclusion}

\label{sec:discuss}

The general properties of the GRB\,990705 host have been summarized in
Table\,1.  According to its morphology, star formation activity and
absolute luminosity, we find that it is typical of the (disk)
galaxies in the field at similar redshifts.

Taking account of the cumulative surface density distribution of
sources with R\,$\leq$\,22.8, \citet{Masetti00} had estimated a
probability of only 0.006 for the burst and the underlying
galaxy being hazardously superimposed on the sky by projection effect,
and had thus suggested a secure identification of this galaxy with the
host of GRB\,990705. Our spectroscopic observations reveal that
the redshift of the spiral is consistent with the one derived by
\citet{Amati00} for the GRB itself, providing further convincing
evidence for a true association between the two.  
Among the current sample of GRB host galaxies, the host of GRB\,990705
has been so far the only case clearly identified with a large
disk-dominated spiral structure at high redshift, the others being
classified either as compact, irregular or interacting systems
\citep[see e.g., Fig.\,2 of][]{Bloom02a}. With an absolute magnitude
M$_B\,\approx$~--21.75 (H$_0$=65, $\Omega_m$=0.3,
$\Omega_{\lambda}$=0.7), it lies furthermore within the brightest
sources of the GRB host sample which is mostly characterized by
sub-luminous systems.  

With a larger sample of GRB hosts, the redshift-dependant proportion
of large disks similar to the host of GRB\,990705 relative to
sub-luminous blue galaxies could provide indications on the fraction
of star formation taking place in massive spirals and thus inform us
of the cosmological evolution of the disk-dominated systems. This
perspective appears promising since such massive and spiral
objects are believed to be responsible for an important fraction of
the Extragalactic Infrared Background \citep{Rigopoulou02}. They could
thus harbor star-forming regions enshrouded in dusty environments
which are not sampled by the blue faint galaxy population.

We finally stress on the remarkable result obtained by \citet{Amati00}
who derived a reliable estimation of the burst redshift interpreting a
transient edge observed in the GRB X-ray spectrum as an iron
absorption at z=0.86\,+/--\,0.17.  
They showed that
intrinsic GRB properties, such as the
redshift and the physical conditions of the GRB--surrounding medium,
 can be derived from the burst detection itself, without the need
of any afterglow to be detected and followed-up.

Eventhough GRB\,990705 is the only one burst
in which such a transient edge has been observed so far, which raises
the question whether particular ionizing states of the circum-burst
environment are required to detect these absorptions,
this burst lies among the brightest GRBs ever detected with the
Beppo-SAX satellite \citep{Amati00}.
This suggests that these transient edges could be a more common
feature of GRB spectra.
Future satellites equipped with more sensitive X-ray detectors, 
such as the ECLAIRs experiment \citep{Barret02},
could be entirely dedicated for studying the GRB
prompt emission, and may provide a
systematic detection of these absorption lines.
Larger samples of GRB redshifts could be derived, an
achievement indeed required to estimate the star formation history
in the Universe from the GRB occurence rate. 

Furthermore, compelling
key results could be obtained towards the class of short GRBs or
specific sub-classes of long GRBs such as the so-called dark
bursts.  The latter,
exhibiting X-ray and radio afterglows without any detected optical
counterparts, could pinpoint not only GRBs with optical
afterglows either locally absorbed by dust or characterized by steep
and rapid decays with time, but also very high redshift GRBs whose
optical emission may be suppressed by the Gunn-Peterson HI trough
along their line of sight. The use of transient features in X-ray
spectra to derive GRB redshifts could thus provide a new approach to
probe very distant GRBs in the early Universe.
In the case of short GRBs, their distance scale and physical origin are
simply still unknown since no detailed follow-up of their afterglows has been
possible so far \citep[but see][]{Castro-Tirado02}. The clues of 
their formation mechanism directly observed in the GRB prompt emission would
 undoubtly
improve our current understanding of these particular events.

\begin{acknowledgements} 
This work largely benefited from the input of  public HST data taken as
part of the ``Survey of the Host Galaxies of Gamma-Ray Bursts'' by
\citet{Holland00a}. We wish to thank especially J.\,Greiner for his
careful reading of the manuscript, as well as F.\,Masset, P.\,Goldoni
and F.\,Daigne for fruitful discussions on this work.  
We also acknowledge the referee for his/her useful comments.
We made extensive
use of publicly available software with material credited to STScI and
prepared for NASA under Contract NAS5-26555. This work was partially
supported by CONICET/Argentina and Fundacion Antorchas.
\end{acknowledgements}

\begin{deluxetable}{lr}
\singlespace
\tablecolumns{2} 
\tablewidth{0in}
\tablecaption{Properties of the GRB\,990705 host galaxy\label{tab}}
\tabletypesize{\footnotesize}
\tablehead{
\colhead{Parameter} & \colhead{Measure}}  

\startdata
$\alpha$ (J2000.0) (host nucleus)     & 05$^h$09$^m$54\,\fs8    \\
$\delta$ (J2000.0) (host nucleus)     & -72$\degr$07$'$54$''$   \\
Foreground extinction E(B--V)	    &  0.12 \\
Spectroscopic redshift		    & 0.8424\,+/--\,0.0002    \\
CL aperture (dered.) magnitude ($\lambda_o$\,=\,5835\,\AA)    & 22.45\,+/--\,0.10       \\
LP aperture (dered.) magnitude ($\lambda_o$\,=\,7208\,\AA)    & 22.00\,+/--\,0.10 \\
Half-light radius  R$_{\,0.5}$    & $\sim$\,7.5\,kpc \\
Central surface brightness $\mu_0$(B$_{AB}$)  & $\sim$~20.8 \\
Absolute M$_B$ magnitude  (H$_0$=65, $\Omega_m$=0.3,$\Omega_{\lambda}$=0.7)         & $\approx$~--21.75 \\
{\it UV-unobscured} Star Formation Rate    & $\sim$~5--8~M$_{\odot}$\,yr$^{-1}$
\enddata
\end{deluxetable}

\end{document}